\documentclass{ws-ijmpc}

\begin{document}

\markboth{A. Chmiel and J. A. Ho{\l}yst }
{Flow of emotional messages in artificial social networks.}

\catchline{}{}{}{}{}

\title{FLOW OF EMOTIONAL MESSAGES IN ARTIFICIAL SOCIAL NETWORKS
}

\author{ANNA CHMIEL}

\address{
Faculty of Physics\\ Center of Excellence
for Complex Systems Research\\ Warsaw University of Technology,
Koszykowa 75\\PL-00-662 Warsaw, Poland\\
anka@if.pw.edu.pl}

\author{JANUSZ A. HO{\L}YST}

\address{Faculty of Physics\\ Center of Excellence
for Complex Systems Research\\ Warsaw University of Technology,
Koszykowa 75\\PL-00-662 Warsaw, Poland\\
jholyst@if.pw.edu.pl}

\maketitle

\begin{history}
\received{Day Month Year}
\revised{Day Month Year}
\end{history}

\begin{abstract}

Models of message flows in an artificial group of users communicating via the Internet  are introduced and investigated using numerical simulations. We assumed that  messages possess an emotional character with a positive valence   and that the  willingness to send  the next affective  message to a given person increases with the number of messages received  from this person. As a result, the weights of links between group  members evolve over time. Memory effects are introduced, taking into account that the preferential selection of message receivers depends on the communication intensity  during the recent period only. We also model the phenomenon of secondary social sharing when the reception  of  an emotional e-mail triggers the distribution of several  emotional e-mails to other people.

\end{abstract}


\section{Introduction}

The World Wide Web (WWW) is the location  of various {\it human actions} that are of interest to many physicists because of the  plethora of available data \cite{Huberman,ksps,ksD,ben,chinren,Bosa,model,return,V,Ha} and the complexity of phenomena taking place in techno-social networks. An example  is the bursty nature of  human activities in cyberspace (e-mails, web-browsing)  considered by Barab\'{a}si  \cite{natureB,Bpre} to be a consequence of decision-based queuing processes. Since  activity patterns in e-communities, e.g.  social groups emerging due to interactions on the Internet \cite{socialcre,offline,group}, are now better understood \cite{pB,ja,Gon}, one can ponder more specific issues related to interactions between  users in the e-world. Until now, there have only been limited results regarding the  influence of emotions on the structures of e-communities. This  issue is specific in the sense that  communication via the Internet is different than meetings in the real world. However, emotions are also expressed in e-mails and even more in anonymous comments on blogs or in Internet discussions. What are emotions? There is no agreement among psychologits about  a common definition of this phenomenon. Here, we shall understand it as follows: 
"Emotions are caused by information processing, so called appraisals, that relates internal or external events to personal relevance and implications, taking into account whether the individual can cope with these and how they relate to personal and social norms"\cite{Arvid}.  An  external event for a specific agent  in cyberspace can be as basic as a message  containing information that evokes an emotion because of the  receiver's  personal connection  to that message. It is possible to  measure the emotions  of  individual users  using a specific equipment \cite{A1,A2,A3}. One  can also efficiently detect emotional content in a text \cite{rudy,mike,pang} employing the methods of machine learning.

To  describe  social relationships one can use  network-based models. In the first approximation, the presence of  any social interaction is shown as a link between two group members. Such networks evolve over time as new group members join and  new social interactions occur \cite{BabasiPA,BS}.  Evolving  {\it weighted} networks are a natural extension of unweighted ones \cite{tele,prltele}. Weight expresses the strength of a social link that can be measured, for example, by the frequency of existing mutual contacts \cite {fri}. It is  worth  stressing that  weight distributions in  social networks \cite{tele} do not always follow a power law, as might be expected  from  the model   developed by  Barrat {\it at al.} \cite{EW}.

Several models of  weighted social network have been studied in the literature.  Results of the model by Kumpula {\it at al.} \cite{prltele}, which assumes a fixed number of users and  evolving  values of weights between them, are in agreement with data about community structures in telephone networks.  Another example of successful  modelling of  social interactions is the  paper by Singer {\it at al.}  \cite{fri}, where  the authors define a possible friendship  as a function of the total number of contacts between the agents and obtain a degree distribution that is in agreement  with the  data collected from social studies about friendship networks in schools. Although the outcomes from  both models are consistent  with collected data, none have dealt with weight distributions in social networks. 

We believe that link weights  are crucial for communities in cyberspace since it is easy to send a single e-mail to a person and only observations of frequent e-mail exchange reveal the significance of real social relations.  Their dynamics are obviously driven by affective phenomena that are introduced in our approach in a simplified way.

\section{Network Structure}
We construct an evolving, directed, weighted network of agents in an artificial community  where  weight $w_{ij}(t)$ is the number of  messages that person $j$ has already received from person $i$ up to the time moment $t$. Generally there is $w_{ij}(t)\neq w_{ji}(t)$ what means that a link form $i$ do $j$ can have a different weight than a link from $j$ to $i$.  Initially, we start with a fully connected network of $N$ agents. The number of agents is fixed in time. We shall try to model   flows of {\it positive} emotional messages among the agents; however, our model can be  generalized to cases where both positive and negative emotional messages are communicated. The initial condition is a fully connected network of agents where every link possesses the same weight, $w_{ij}(0)=1$.  This means that at the very beginning, each group member sent a polite message, such a {\it Hello Partner}, to every other group member; thus, the initial weight probability distribution is $P(w)=\delta_{w,1}$. During the evolution process weights become heterogeneous.

\section{Models of  emotional  message transfer in a group}
To model a process of  emotional message transfer in our e-community, we use several variants of the updating rules, starting with the simplest and most trivial cases and then moving to more complex and more realistic solutions. Our main goal is to determine the most important characteristics of this toy model, which will eventually allow us to look for similarities to real-world data in the future.

\subsection{Model 0 }

Model 0 is a trivial case where in every time step we randomly choose a sender, $i$, of an emotional message and then randomly find the recipient, $j$. This processes an increase in the number of transferred messages   between nodes $i$ and $j$ so that $w_{ij}(t+1)=w_{ij}(t)+1$. In such an approach, the fact that messages contain emotional content that can influence agents' actions has no impact.  This simple updating procedure gives the Poisson  distribution of weights $P(w)=\frac{\lambda^w e^{-\lambda}}{w!}$, where $\lambda=Tp$,$T$ is the simulation time, and  $p= \frac{1}{N(N-1)}$.

\subsection{Model I -- with infinite memory}

Unlike the previously described version, {\bf Model I } is equipped with a memory of  emotions that were communicated between agents.   We randomly find the sender (agent $i$) and then we choose a recipient for his emotional message (agent $j$) using a preferential rule. The  agent decides to whom he wants to send a message, taking into account the complete  history of communications with other agents.
 The selection probability is proportional to the number of past messages  transferred between nodes $j$ and $i$. 
 The reason for this rule is that the agent $i$  makes a stronger (emotional) contact with the group members who  have already sent him many messages. Here, we stress the fact that, for simplicity, all messages express  only positive emotions. In the continuous time approximation, our rule of communications will lead to the following equation for changes of the weights $w_{ij}(t)$:
\begin{equation}
\frac{\partial{w_{ij}(t)}}{\partial{t}}=\frac{w_{ji}(t)}{s^{in}_i(t)} \frac{1}{N}.
\label{pw_sm}
\end{equation}
Here, $s^{in}_i(t)$ is the temporal incoming strength of node $i$  defined  as $s^{in}_i(t) = \sum_{j}{w_{ji}(t)}$. 
Numerical results show that with good agreement we can say $w_{ij} \approx w_{ji}$ and 
\begin{equation}
\frac{\partial{w_{ij}(t)}}{\partial{t}}=\frac{w_{ij}(t)}{s^{out}_i(t)} \frac{1}{N}
\label{pw_sm1}
\end{equation}
where $s_{i}^{out}= \sum_{j}{ w_{ij}}$  can be estimated as $s_{i}^{out}(t) \approx \langle s^{out} \rangle (t)$. However, the mean value of node strength is  $\langle s^{out} \rangle (t)=\frac{N(N-1)+t}{N}= t/N +N-1$. The number of links in our problem is constant, similar to the number of nodes in Barabasi-Albert model B \cite{BabasiPA}, in which the  authors applied the  preferential attachment rule without adding new nodes. In our, case we assume a preferential increase of  directed weights. In the above-mentioned model B, the degree of node is proportional to $t$, i.e. $k_i(t) \sim t$ for $t>N$. In our problem, mean-field-like analysis, i.e. substituting $s_{i}^{out}$ with $t/N +N-1$, and putting it to Eq.(2) leads to the equation:
\begin{equation}\label{pw_sm1}
w_{ij}(t) = \frac{N(N-1)+t}{N(N-1)+1} \approx \exp (\frac{t-1}{N(N-1)}),
\end{equation}
which is correct for large values of $N$ and $t\ll N^2$, where we used the initial condition $w_{ij}(t=1)=1$. The process of weight increase is very slow in the case of large systems.

\begin{figure}[ht]
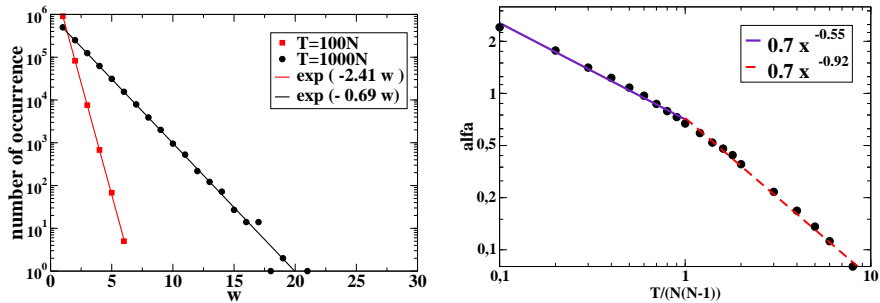

\vskip 0.5cm
\centerline{\begin{tabular}{ c c}
 \epsfig{file=pw1.eps,width=.44\textwidth} & \epsfig{file=alfa_nn.eps,width=.44\textwidth} \\
\end{tabular}}
    \caption{The weight distribution is shown on the left, and the relationship between the parameter $\alpha$ and $\frac{T}{N(N-1)}$ is shown on the right (Model I).}
    \label{wag_sm}
\end{figure}
Observing  the weight distributions $P(w)$  after $T$ steps of simulation,  we found that it follows an exponential behaviour with a characteristic exponent $\alpha$ : 
\begin{equation}\label{pw_sm}
P(w)=A \, \exp(-\alpha \, w).
\end{equation} 
In Fig. \ref{wag_sm}, we present examples of weight distributions for different values of the time of simulation $T$ for networks with $N=1000$. A more detailed analysis of the dependence of exponent $\alpha$ on the total time divided by the system size $\frac{T}{N^2}$ is shown on the right side of  Fig. \ref{wag_sm}. As one can see, the value of $\alpha$ decreases  with $\frac{T}{N^2}$ following a power-law behaviour $\alpha \sim (T/N^2)^{-\beta}$, where two different regions of scaling can be distinguished.

\subsection{Model II --- with temporary memory}
We develop {\bf Model I} by introducing the concept of memory length. This approach allows us to consider a more  realistic situation in which users  forget  very remote past events and make their decisions based only on  the last transfer of messages. The updating procedure is almost identical to that in  {\bf Model I}, except  that the selection probability is proportional to the number of past emotional messages  transferred between nodes $j$ and $i$ in the last $cN$ steps:
\begin{equation}\label{pw_sm}
\frac{\partial{w_{ij}(t)}}{\partial{t}}=\frac{w_{ji}(t)-w_{ji}(t-cN)}{s^{in}_i(t)-s^{in}_i(t-cN)}\frac{1}{N}.
\end{equation}

\begin{figure}
\vskip 0.5cm
 \centerline{\epsfig{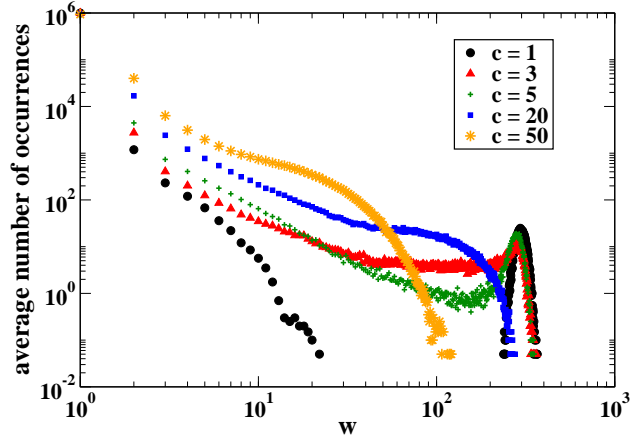}}
    \caption{Weight distribution for $T=300N$ and $N=1000$; 20 trials (Model II).  }
        \label{fig_wagi_C}
\end{figure}
\begin{figure}
\vskip 0.6cm
 \centerline{\epsfig{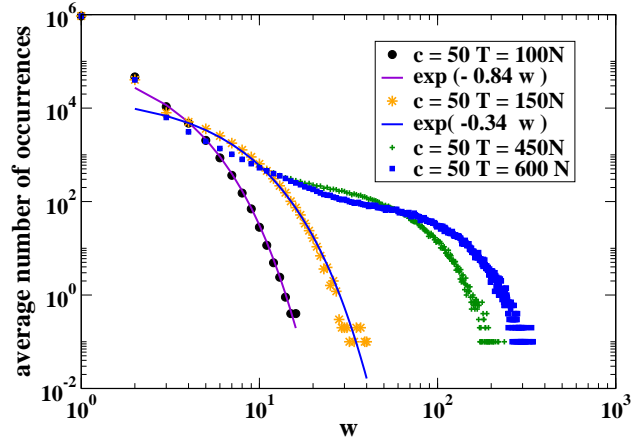}}
    \caption{Weight distribution, for $N=1000$ and $T=300N$; 20 trials (Model II). }
        \label{fig_d}
\end{figure}

 We observe how the weight distribution changes with parameters $T$ and $c$.
In the case of $c<3$ for small values of $w_{ij}$ the distribution $P(w_{ij})$  decreases while for larger $w_{ij}$, one can observe a Gaussian-like peak with the mean  value equal to $T/N$ (see Fig. \ref{fig_wagi_C}). The value $c=3$ is a transition point; here, the two parts of the distribution merge, i.e. the Gaussian peak is absorbed by the decreasing part. For $c>5$, there are once again traces of the Gaussian curve. For larger values of $c$ (e.g. $c=20$ and $c=50$), the Gaussian part is completely invisible.   
To measure the evolution of weight distribution  with changing memory length, we analyse the standard deviation defined as 

\begin{equation}\label{pw_sm}
\sigma=\sqrt{\frac{1}{N(N-1)} \sum_{ij}(w_{ij}-\langle w \rangle)^2},
\end{equation}
where $\langle w \rangle=\frac{1}{N(N-1)}\sum_{ij}w_{ij}$ is the average weight of a link. The analysis confirms our previous observation that the lengths of memory $c=3$ and $c=5$ are transition points; in the standard deviation analysis, those points form  local minimum and maximum, respectively (see Fig. \ref{fig_d}).

\begin{figure}
\vskip 0.5cm
 \centerline{\epsfig{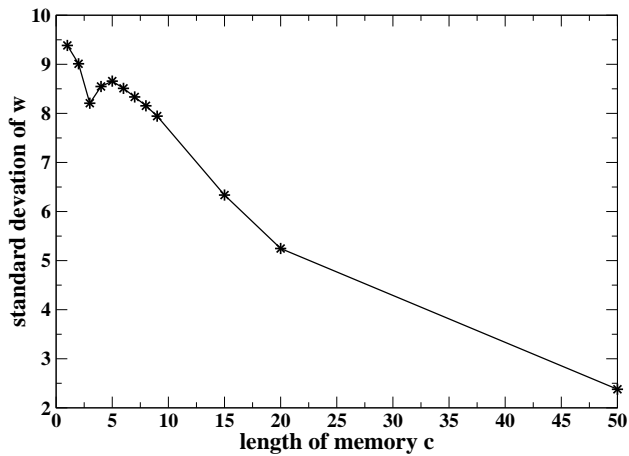}}
    \caption{Standard deviation as a function of the length of memory ($T=300N$, $N=1000$; 20 trials (Model II)).  }
        \label{fig_wagi_sing}
\end{figure}

For larger values of $c$, for example $c=50$, we observed the following behaviour: for $T=100N$, the character of the distribution is close to exponential (without the first point, which is dependent on the initial conditions), while for $T=450N$, a fat tail with exponential cutoff is visible (see Fig. \ref{fig_d}).

\subsection {Model III --- with secondary social sharing }
Secondary social sharing is a phenomenon  widely described  in psychology. According to  the  famous researcher of this problem,  Bernad Rim\'{e} \cite{BR,tBR,BRrew}, {\it"First, it was predicted that being exposed to the social sharing of an emotion is emotion-inducing. Second, it was reasoned that if this holds true, then the listener should later engage in socially sharing with other persons the emotional narrative heard. Thus, a process of `secondary social sharing' was predicted." }
 It was also proved by  Rim\'{e}, \cite{tBR} that the probability of social sharing phenomenon increases with the level of triggering  emotions.
 
\begin{figure}
\vskip 0.5cm
 \centerline{\epsfig{file=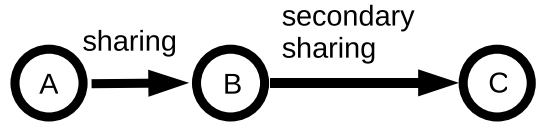,width=.55\columnwidth}}
    \caption{Scheme of secondary social sharing process.  }
        \label{fig_sss}
\end{figure}

\begin{figure}
\vskip 0.5cm
 \centerline{\epsfig{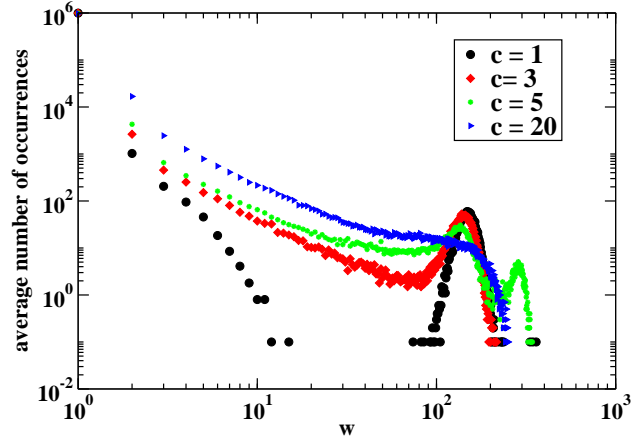}}
    \caption{Weights distribution, for  $N=1000$ and $T=300N$; 10 trials (Model III). }
        \label{fig_soc}
\end{figure}

We try to adapt this phenomenon  in order to extend  our  {\bf Model II}. As the level of emotion is not considered in our models, we assumed that a secondary person  shares  the emotional message with only  one person (see Fig. \ref{fig_sss}). We randomly find a sender of a message and find the receiver using the rule from{  \bf Model II }. This receiver will then send a message to next person, using the same rule. The results obtained for this version of the model (see Fig. \ref{fig_soc}) are similar to Model II (see Fig. \ref{fig_wagi_C})  with temporary memory; we also observe the decreasing part and Gaussian behaviour for the small value of $c$. A new feature  of this distribution is an additional  peak, for $c=5$.

\subsection {Model IV --- chain letter }
\begin{figure}
\vskip 0.5cm
 \centerline{\epsfig{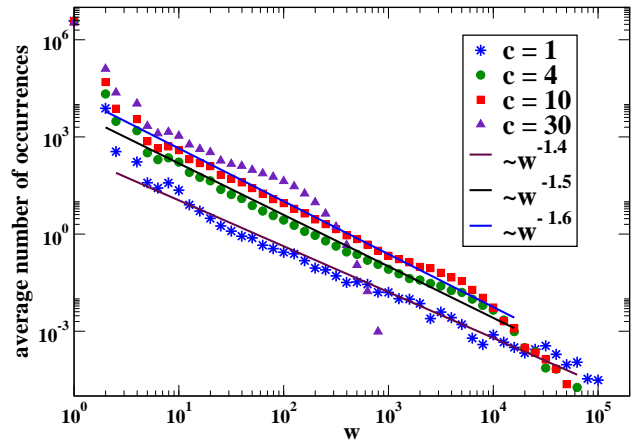}}
    \caption{Weights distribution, for $N=1000$ and $T=300N$; 10 trials (Model IV). }
        \label{fig_wagi_ss}
\end{figure}

In this version, we assume that  one user can send several emotional messages and one of their recipients becomes the sender of the next emotional message. The basic preferential rule with memory is still the same. We randomly select a user and randomly find the number of users to whom this agent sends a message. One of the  recipients will send the message to a random number of users, which creates a chain of messages.
The  weighted distribution obtained from this version of the  model is presented in  Fig. \ref{fig_wagi_ss}. For $c < 30$, we can find a power-law scaling in the central regime (the first point is due to the  influence of the initial condition and at the end one, observes an    exponential cutoff). This behaviour is qualitatively different from those in all previous cases. The introduction  of the chain  rule plays a pivotal role here. One can see the similarity to the random  walk problem in a weighted network \cite{ja}.

\section{CONCLUSIONS}
We analysed   several  models of communicating  emotional messages  in  artificial e-communities that are described by directed weighted networks, where the weights/strengths  of links correspond to the total number of messages sent from one agent to another.  Model 0 assumes a random evolution of link strengths, $w_{ij}(t)$, and as a result, the distribution  $P(w)$ is Poissonian. In Model I, we  assumed that  messages possess an emotional character with a positive valence  and that the  willingness to send  an  affective  message to a given person increases linearly with the number of messages already received  from this person. As a result, we obtain an exponential behaviour of  $P(w)$,  where the characteristic   exponent depends in a unique way on   $T/N^2$  (N is a network size and $T$  is a network age). Introduction of  a limited memory length $c$ into  Model II significantly changes the distributions $P(w)$  that become the sum of the  monotonically  decaying part and a Gaussian peak. For larger values of the memory window $c$, the peak merges with the monotonic part. The effects of  secondary social sharing phenomena were considered in Model III; however, the results were similar to those of  Model II. In Model IV, we assumed a   chain rule form of e-mail transmissions that  resulted  in the  power-law behaviour of $P(w)$.
The assumption of the finite memory length in Model II can be specific   for some e-services where  only  limited temporal  access to previously exchanged messages exists.

\section*{Acknowledgments}

We acknowledge the useful discussion about emotion phenomena with Arvid Kappas, Dennis Kuster, Elane Tsankova and Mathias Theunis. All remaining errors are ours. We are thankful to Julian Sienkiewicz for his  critical reading of this manuscript. The work was supported by EU FP7 ICT Project  {\it Collective Emotions in Cyberspace - CYBEREMOTIONS}, European COST Action MP0801 {\it Physics of Competition and Conflicts} and Polish Ministry of Science Grant 1029/7.PR UE/2009/7 and Grant  578/N-COST/2009/0.


\end{document}